\documentclass[12pt]{article}
\usepackage{authblk}
\usepackage{natbib}
\usepackage{amsmath, amsfonts,mathrsfs}
\usepackage{color}
\usepackage{rotating}
\usepackage{lscape}
\usepackage{graphicx,multirow}
\usepackage{bm,url}
\usepackage{titling,subfigure}
\usepackage{float}
\usepackage{longtable}

\usepackage{amsmath}
\usepackage{caption} 
\newtheorem{lemma}{\noindent\mbox{Lemma}}
\newtheorem{theorem}{\noindent\mbox{Theorem}}
\newtheorem{corollary}{\noindent\mbox{Corollary}}
\newtheorem{proposition}{\noindent\mbox{Proposition}}
\newtheorem{condition}{\noindent\mbox{M}}

\def\be{\begin{equation}}
\def\ee{\end{equation}}
\def\bea{\begin{eqnarray}}
\def\eea{\end{eqnarray}}
\def\bd{\begin{displaymath}}
\def\ed{\end{displaymath}}
\def\bda{\begin{eqnarray*}}
\def\eda{\end{eqnarray*}}
\def\bsm{\begin{small}}
\def\esm{\end{small}}

\def\ha1{\hat \beta_1}

\def\bb0{\delta_\beta}

\def\bsc{\begin{scriptsize}}
\def\esc{\end{scriptsize}}

\newcommand{\blind}{1}

\begin{document}

\def\spacingset#1{\renewcommand{\baselinestretch}%
{#1}\small\normalsize} \spacingset{1}

\if1\blind
{
  \title{\bf Finite Sample t-Tests for High-Dimensional Means}
  \author{Jun Li \hspace{.2cm}\\
    Department of Mathematical Sciences, Kent State University}
  \maketitle
} \fi

\if0\blind
{
  \bigskip
  \bigskip
  \bigskip
  \begin{center}
    {\LARGE\bf Finite Sample t-Tests for High-Dimensional Means}
\end{center}
  \medskip
} \fi

%%%%%%%%%%%%%%%%%%%%%%%%%%% Abstract %%%%%%%%%%%%%%%%%%%%%%%%%%%%%%%%

\bigskip
\begin{abstract}
Size distortion can occur if an asymptotic testing procedure requiring diverging sample sizes,  is implemented to data with very small sample sizes. %It can be even more challenging to maintain an accurate size when data are high-dimensional. 
In this paper, we consider one-sample and two-sample tests for  mean vectors when data are high-dimensional but sample sizes are very small. We establish asymptotic $t$-distributions of one-sample and two-sample $U$-statistics, which only require data dimensionality to diverge but sample sizes to be fixed and no less than $3$.  Simulation studies confirm the theoretical results that the proposed tests maintain accurate empirical sizes for a wide range of sample sizes and data dimensionalities. We apply the proposed tests to an fMRI dataset to demonstrate the practical implementation of the methods.        
 
\end{abstract}

\noindent%
\vfill
%%%%%%%%%%%%%%%%%%%%%%%%%% INTRO %%%%%%%%%%%%%%%%%%%%%%%%%%%%%%%
\newpage
\spacingset{1.9} % DON'T change the spacing!
\section{Introduction}
\label{sec:intro}

%\setcounter{section}{1} \setcounter{equation}{0}
%\section*{\large 1. \bf Introduction}

Testing for population means is a classical problem in statistics. In the univariate case, the Student's $t$-test can be applied when the sample mean follows a normal distribution, sample variance follows a $\chi^2$ distribution and the sample mean and sample variance are independent.  
In the traditional multivariate setting, Hotelling's $T^2$ test (Hotelling, 1931) can be applied when dimension $p$ is fixed and dimension $p$ and sample size $n$ satisfy the relation $p \le n-1$.

With the explosive development of high-throughput technologies, high-dimensional data characterized by the ``large $p$ and small $n$'' situation are widely observed.  When $p>n-1$, the Hotelling's $T^2$ test %encounters an unsatisfied performance
becomes infeasible due to singularity of the sample variance. Even when $p$ is close to $n-1$, the Hotelling's test loses its power as revealed by Bai and Saranadasa (1996).
Many approaches have been proposed to modify the Hotelling's $T^2$ test for high-dimensional data. Some were constructed to discard or stabilize the inverse of the sample variance. Examples include Bai and Saranadasa (1996), Srivastava and Du (2008), and Chen and Qin (2010). %More can be seen in Wang, Peng and Li (2015), Chen et al. (2011), and Li et al. (2016). %that is utilized to incorporate data dependence, because it is not a consistent estimator under high dimensionality. However, simply ignoring dependence in the test statistics will lose a large part of the mutual information contained in the data.
%A drawback of
Some were proposed to reduce the noise contributed by non-signal bearing components for sparse signal detection. Examples include the maximum type test proposed in Cai, Liu and Xia (2014) and the thresholding tests in Hall and Jin (2010), Zhong, Chen and Xu (2013) and Chen, Li and Zhong (2019). 
%Nevertheless, the implementation of the thresholding tests as well as the maximum test requires strict sparsity structure assumptions on the covariance or precision matrix, %such as the polynomial off-diagonal decay in Hall and Jin (2010). In real applications, these assumptions
%which may not be satisfied in real applications.
Others were proposed to project the classical Hotelling's $T^2$ statistic to a low-dimensional space. 
Examples include Thulin (2014), and Srivastava, Li and Ruppert (2016). 
All these testing procedures were established by requiring dimensionality and sample sizes to diverge even though dimensionality can be much larger than sample sizes.  

%Even though they are designed for the high-dimensional data with a ``large $p$, small $n$'' situation, the asymptotic results rely on the assumption that both dimensionality and sample sizes diverge. In many preclinical studies, high dimensional data with very small sample sizes often occur  due to ethical, financial, and feasibility reasons. If we apply aforementioned tests to such data with very small sample sizes, the size or the probability of type I error may not be properly controlled.  This can lead to either being aggressive or conservative to reject the null hypothesis.    

In many biological and financial studies, high dimensional data with very small sample sizes often occur due to ethical and feasibility reasons. As demonstrated by the simulation studies in Section 4, when sample sizes are very small, implementing a testing procedure established by requiring diverging sample sizes may not control the type-I error and thus wrongly reject the null hypothesis with a probability higher or lower than a preselected nominal significance level.    
To take into account the small sample effect, we establish the asymptotic normality of a standardized one-sample $U$-statistic by only requiring data dimensionality to diverge to infinity. In practice the standard deviation of the $U$-statistic is unknown and cannot be consistently estimated when sample size is small. By analogy with the univariate Student's $t$ statistic, we  propose an estimator for the variance of the $U$-statistic, which is shown to be asymptotically $\chi^2$ distributed and independent to the $U$-statistic under the null hypothesis. The result enables us to establish the asymptotic $t$-distribution of the  $U$-statistic standardized by the sample standard deviation. The test only requires data dimensionality to diverge but the sample sizes to be fixed and no less than $3$. Moreover, it is nonparametric without assuming Gaussian distribution of data. We further extend the asymptotic results to the two-sample testing problem. %Numerical studies show that the proposed testing procedures maintain the empirical sizes very accurate to the nominal significance level for a wide range of sample sizes.  

The rest of the paper is organized as follows. Section 2 introduces the one-sample $U$ statistic and establish its asymptotic $t$-distribution.  Extension to the two-sample problem is provided in Section 3. 
Simulation and case studies are presented in Sections 4 and 5. %Section 6 concludes the paper with discussion. 
Technical proofs of theorems are relegated to Appendix. %Additional simulation results are included in a supplementary material.

%\setcounter{section}{2} \setcounter{equation}{0}
%\section*{\large 2. \bf One-sample test}
\section{One-sample test}
\label{sec:one-sample}

Let $X_1, \cdots, X_{n}$ be independent and identically distributed $p$-dimensional random vectors with mean $\mu=\mbox{E}(X_i)$ and covariance matrix $\Sigma=\mbox{Var}(X_i)$.  The hypotheses we are interested in are
\begin{eqnarray}
H_0: \mu=0 \quad \mbox{versus} \quad H_1: \mu \ne 0. \label{Hypo1}
\end{eqnarray}
%which are general because Any other test about whether $\mu$ equals a nonzero vector $\mu_0$ can be converted into the above hypotheses by subtracting each observation with $\mu_0$.  

To test the hypotheses, we consider the following $U$-statistic
\begin{eqnarray}
U_n=\frac{2}{n(n-1)}\sum_{i<j}^{n}X_i^{\prime}X_j, \label{stat1}
\end{eqnarray}
which is the one-sample version of the $U$-statistic proposed in Chen and Qin (2010). 

Like Bai and Saranadasa (1996) and Chen and Qin (2010), we model the sequence of $p$-dimensional random vectors $\{X_i, 1\le i \le n\}$ by a linear high-dimensional time series
\be
X_{i}=\mu+\Gamma Z_i \qquad \mbox{for} \quad i=1, \cdots, n,   \label{model}
\ee
where $\mu$ is the $p$-dimensional population mean, $\Gamma$ is a $p \times q$ matrix with $q\ge p$ satisfying $\Gamma \Gamma^{\prime}=\Sigma$, and $Z_i=(z_{i1}, \cdots, z_{iq})^{\prime}$ so that $\{z_{il}\}_{l=1}^q$ are mutually independent and satisfy $\mbox{E}(z_{il})=0$, $\mbox{Var}(z_{il})=1$ and $\mbox{E}(z_{il}^4)=3+\eta$ for some finite constant $\eta$.

As discussed in Chen and Qin (2010),  the model (\ref{model}) considers data beyond the commonly assumed Gaussian distribution, so that an established testing procedure is nonparametric. It is worth mentioning that different from Chen and Qin (2010), we assume all the components of $Z_i$ are mutually independent rather than pseudo-independent. As shown in the proof of Theorem 1, such a requirement allows us to drop the requirement of diverging sample size $n$ when applying the Martingale central limit theorem to establish the asymptotic distribution of the $U$ statistic. In addition, we assume the following condition for the covariance matrix $\Sigma$.

\medskip
(C1). As $p \to \infty$, $\mbox{tr}(\Sigma^4)=o\{\mbox{tr}^2(\Sigma^2) \}$. 

The condition is automatically satisfied if all the eigenvalues of $\Sigma$ are bounded, but it also allows partial eigenvalues to be unbounded (see Chen and Qin, 2010 for detailed discussion). An advantage of (C1) is that it does not impose any explicit growth rate on the dimension $p$.   

We establish the asymptotic normality of the $U$-statistic as follows. 

\medskip
{\bf Theorem 1.} Assume the model (\ref{model}) and the condition (C1). For any finite sample size $n \ge 2$, 
\[
\frac{U_n-\mu^{\prime}\mu}{\sigma_n} \xrightarrow{d} N(0, 1) \quad \mbox{as} \quad p \to \infty,
\]
where 
\[
\sigma_n^2=\frac{2}{n(n-1)}\mbox{tr}(\Sigma^2)+\frac{4}{n}\mu^{\prime}\Sigma \mu.
\]
Especially, under $H_0$ of (\ref{Hypo1}),
\[
\frac{U_n}{\sigma_{n,0}} \xrightarrow{d} N(0, 1) \quad \mbox{as} \quad p \to \infty,
\]
where 
\[
\sigma_{n,0}^2=\frac{2}{n(n-1)}\mbox{tr}(\Sigma^2).
\]

\medskip

%While the asymptotic normality of $U_n$ was also established in Chen and Qin (2010), they required both data dimensionality $p$ and sample size $n$ diverge to infinity. Unlike Chen and Qin (2010), we only require $p$ to diverge but $n$ to be fixed. 
To implement a testing procedure, we need to estimate the unknown $\mbox{tr}(\Sigma^2)$. 
If the sample size $n$ diverges to infinity, $\mbox{tr}(\Sigma^2)$ can be estimated consistently by some unbiased estimators such as the $U$ statistic in Li and Chen (2012). Slutsky's theorem then shows that the asymptotic normality of $U_n$ with the estimated $\sigma_{n,0}$ still holds. However, we consider the sample size to be small and a consistent estimator of $\mbox{tr}(\Sigma^2)$ is not achievable.  By analogy with the Student's $t$-statistic,  it can be shown that we only need to construct an estimator which is $\chi^2$ distributed and asymptotically independent of the statistic $U_n$. 
Let $\{X_i^{\prime} X_j, i<j\}_{i,j=1}^n$ be a set of $n(n-1)/2$ random variables. From the proof of Theorem 1, the random variables in $\{X_i^{\prime} X_j, i<j\}_{i,j=1}^n$ are asymptotically mutually independent and each component has the variance $\mbox{tr}(\Sigma^2)$ under the null hypothesis.   
Based on $\{X_i^{\prime} X_j, i<j\}_{i,j=1}^n$, we therefore estimate the unknown $\mbox{tr}(\Sigma^2)$ by the sample variance
\begin{eqnarray}
 \widehat{\mbox{tr}({\Sigma}^2)}
= \frac{1}{n(n-1)/2-1} \sum_{i < j}^{n} \biggl(X_{i}^{\prime}
X_{j}-\frac{2}{n(n-1)}\sum_{i<j}^{n}X_{i}^{\prime}
X_{j}\biggr)^2. \label{var.est}
\end{eqnarray}
Its asymptotic $\chi^2$-distribution can be established as follows. 

\medskip
{\bf Theorem 2.} Let the degrees of freedom $k=n(n-1)/2-1$. Assume the model (\ref{model}) and the condition (C1). For any finite sample size $n \ge 3$ and under $H_0$ of (\ref{Hypo1}), 
\[
\frac{k \,\, \widehat{\mbox{tr}({\Sigma}^2)}}{\mbox{tr}(\Sigma^2)} \xrightarrow{d} \chi^2(k)  \quad \mbox{as} \quad p \to \infty.
\]
\medskip

After replacing $\mbox{tr}(\Sigma^2)$ by the estimator $ \widehat{\mbox{tr}({\Sigma}^2)}$, we estimate $\sigma^2_{n,0}$ by 
\[
\hat{\sigma}^2_{n,0}=\frac{2}{n(n-1)}  \widehat{\mbox{tr}({\Sigma}^2)}. 
\]

We establish the asymptotic $t$-distribution of $U_n/\hat{\sigma}_{n,0}$ as follows.

\medskip
{\bf Theorem 3.} Assume the same conditions in Theorem 2. For any finite sample size $n \ge 3$ and under $H_0$ of (\ref{Hypo1}), as $p \to \infty$,  
\[
\frac{U_n}{\hat{\sigma}_{n, 0}} \xrightarrow{d} t \quad \mbox{with} \quad k=n(n-1)/2-1 \quad \mbox{degrees of freedom}.
\]

\medskip

\begin{figure}[t!]
\begin{center}
\includegraphics[width=0.32\textwidth,height=0.32\textwidth]{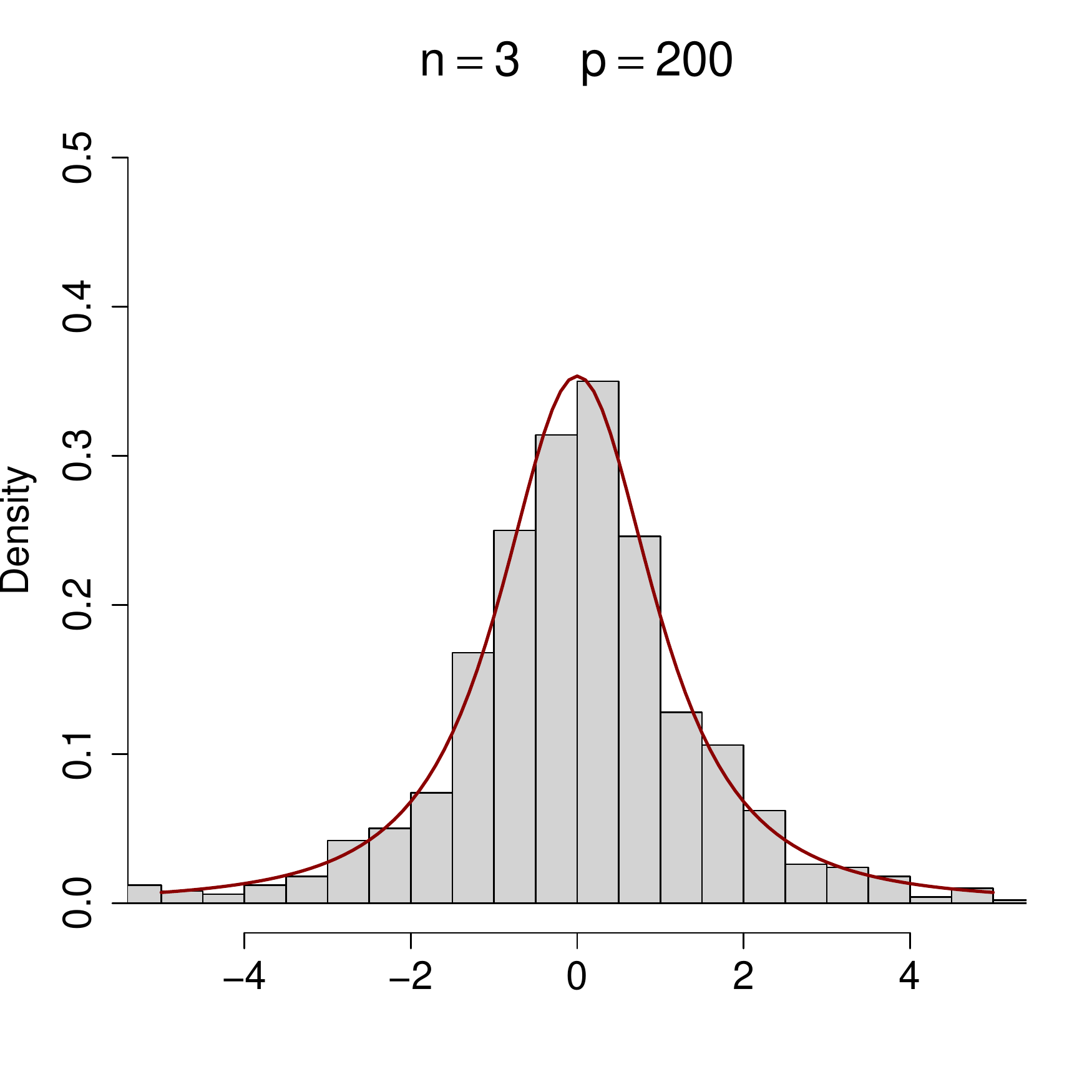}
\includegraphics[width=0.32\textwidth,height=0.32\textwidth]{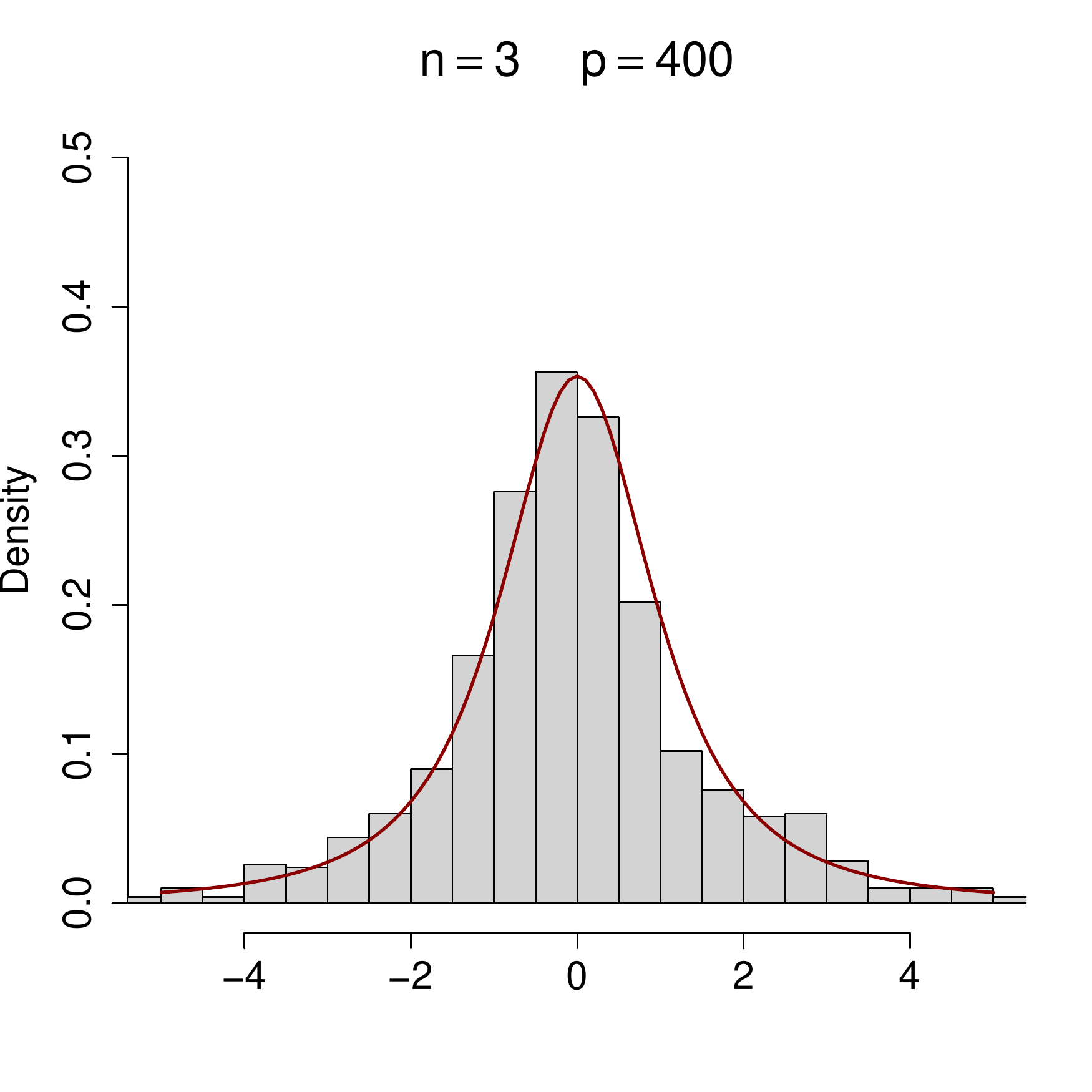}
\includegraphics[width=0.32\textwidth,height=0.32\textwidth]{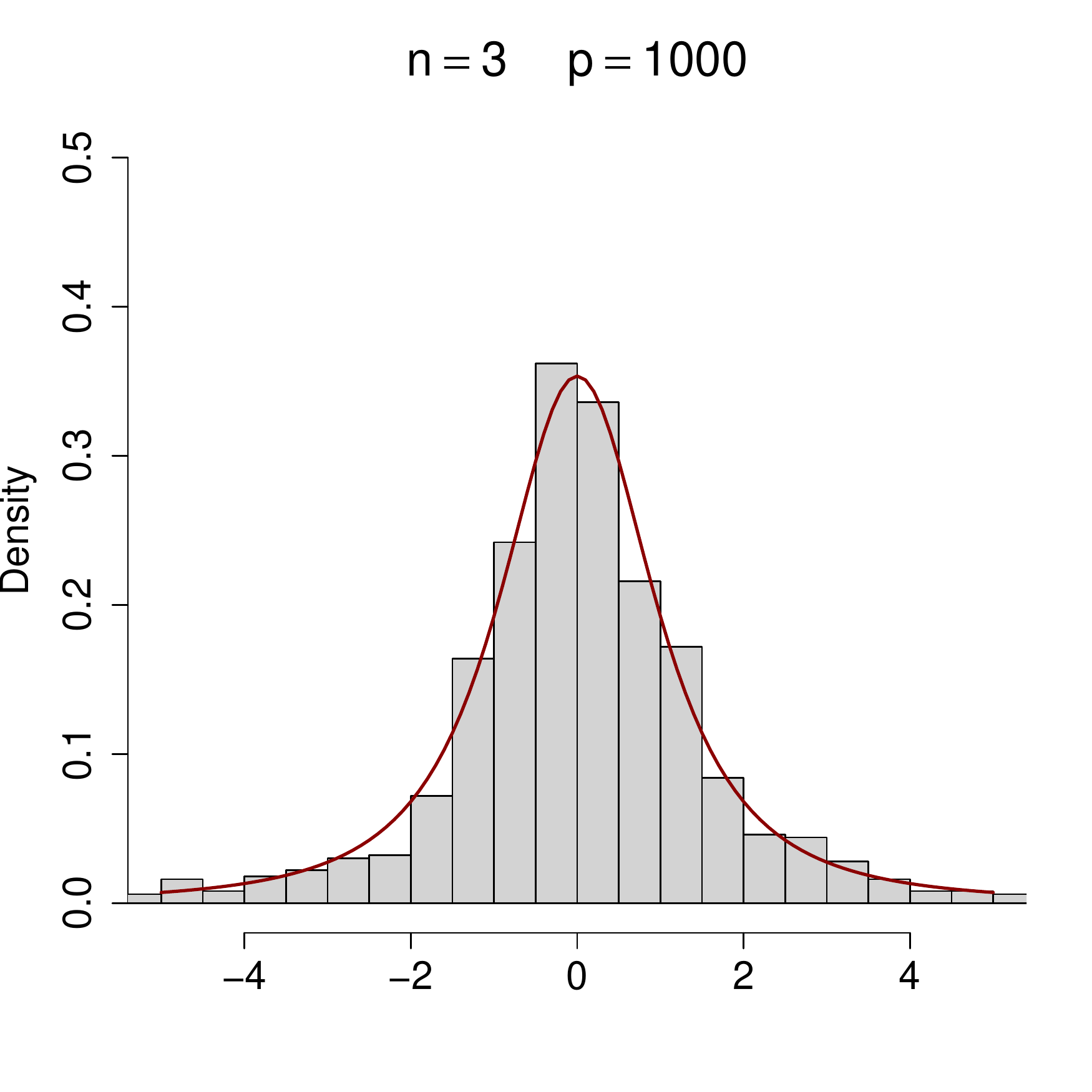}\\
\includegraphics[width=0.32\textwidth,height=0.32\textwidth]{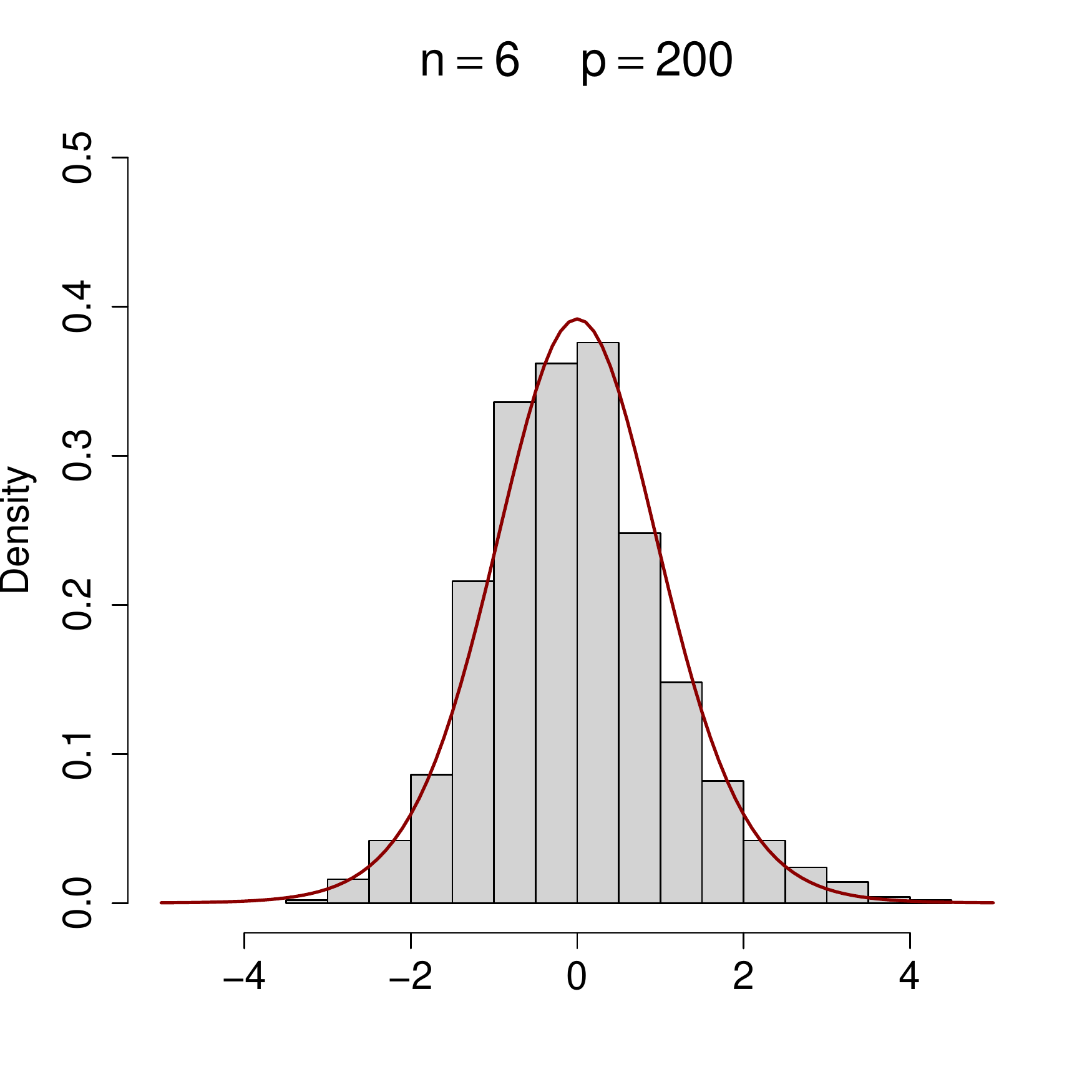}
\includegraphics[width=0.32\textwidth,height=0.32\textwidth]{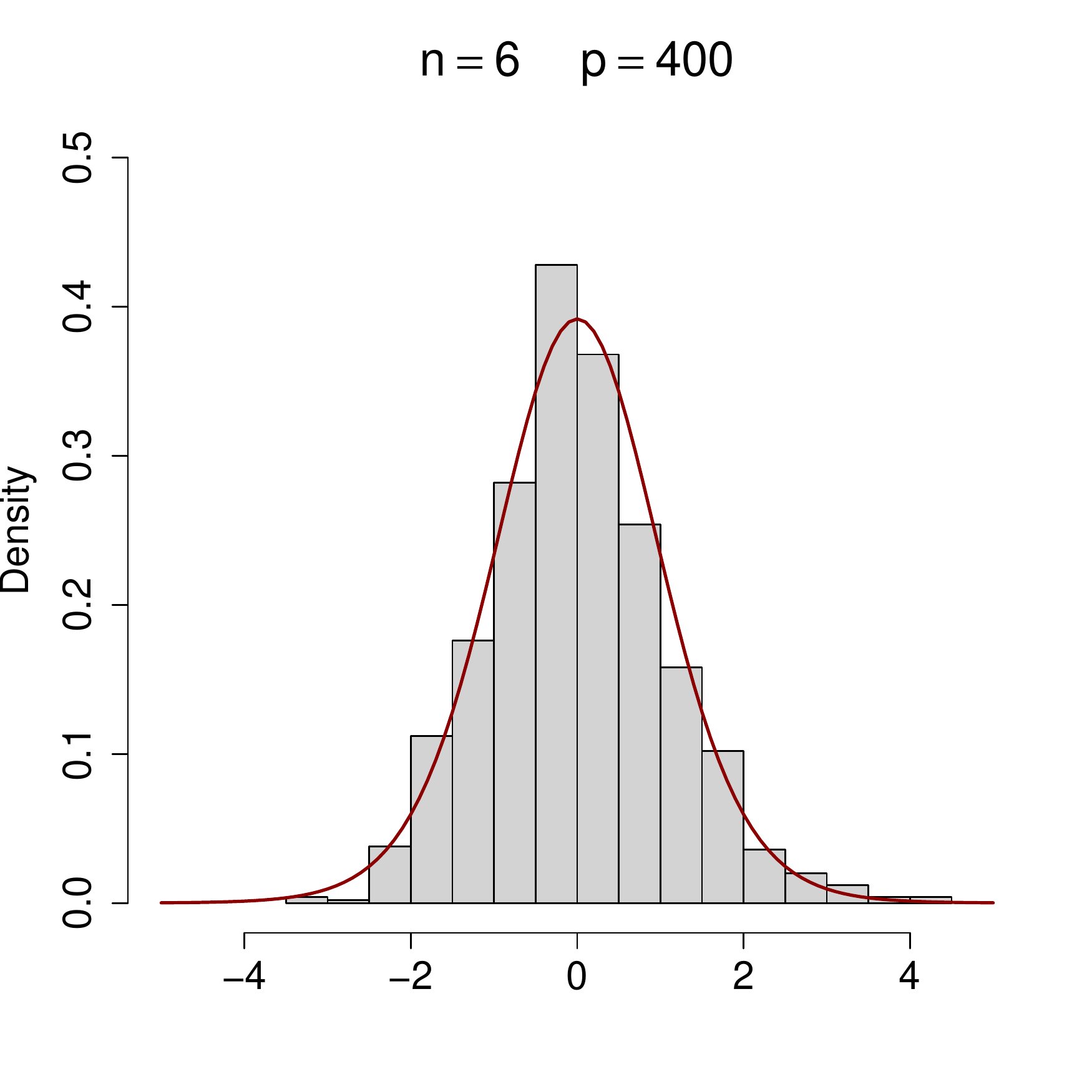}
\includegraphics[width=0.32\textwidth,height=0.32\textwidth]{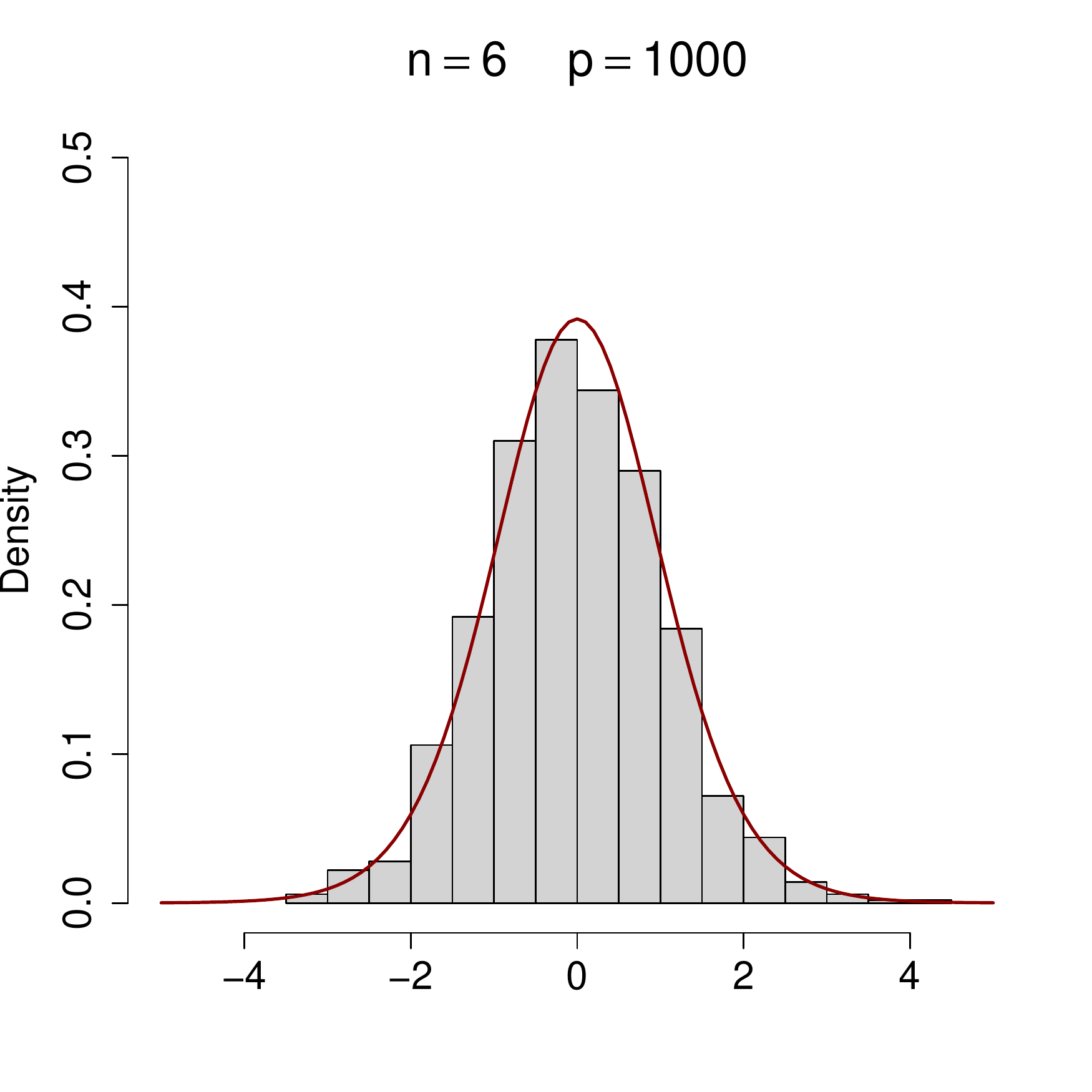}
\caption{Histogram of ${U_n}/{\hat{\sigma}_{n, 0}}$ versus $t$-distribution. In the upper row, the $t$-distribution has the degrees of freedom equal to $2$; In the lower row,  the $t$-distribution has the degrees of freedom equal to $14$. }
\label{fig1-1}
\end{center}
\end{figure}

To put the above result into a visual inspection, we simulate data from $N(0, \Sigma)$ where $\Sigma=(\sigma_{i j})=(0.6^{|i-j|})$. Figure \ref{fig1-1} shows histograms of ${U_n}/{\hat{\sigma}_{n, 0}}$ based on $1000$ iterations with different sample sizes $n$ and data dimensionality $p$. In the upper row, we choose the sample size $n=3$ and the resulted $t$-distribution in Theorem  3 has $2$ degrees of freedom. The empirical distributions of ${U_n}/{\hat{\sigma}_{n, 0}}$ are close to the $t$-curve despite its heavy tails. Similar results can be observed in the lower row where the $t$-distribution has $14$ degrees of freedom and thin tails.

Based on Theorem 3, the proposed test with a nominal $\alpha$ significance level rejects $H_0$ if $U_n/\hat{\sigma}_{n, 0} \ge t_{\alpha}(k)$, where $t_{\alpha}(k)$ is the upper $\alpha$ quantile of $t$-distribution with $k=n(n-1)/2-1$ degrees of freedom. Moreover, the power function of the test when $\mu=\mu_0 \ne 0$ is 
 \begin{eqnarray}
B_1(||\mu_0||^2)&=& \mbox{P}\biggl(\frac{U_n}{\hat{\sigma}_{n, 0}} \ge t_{\alpha}(k)| \mu =\mu_0\biggr)\nonumber\\
& =&1- \mbox{P}\biggl(\frac{U_n-||\mu_0||^2}{\sigma_n} <\frac{\hat{\sigma}_{n, 0}}{\sigma_n} t_{\alpha}(k)-\frac{||\mu_0||^2}{\sigma_n}\quad | \mu =\mu_0\biggr)\nonumber\\
&=&1-\Phi\biggl(\frac{\hat{\sigma}_{n, 0}}{\sigma_n} t_{\alpha}(k)-\frac{\sqrt{n(n-1)}||\mu_0||^2}{\sqrt{2\mbox{tr}(\Sigma^2)+4(n-1)\mu_0^{\prime}\Sigma \mu_0}}\biggr), \quad \mbox{as} \, p \to \infty \nonumber
 \end{eqnarray}
where $||\mu_0||^2=\mu_0^{\prime}\mu_0$, $\sigma_n$ is given in Theorem 1 and $\Phi(\cdot)$ is the cumulative distribution function of the standard normal. 

To see how the power $B_1(||\mu_0||^2)$ evolves with the sum-of-squares signal strength $||\mu_0||^2$, we first derive $$\mbox{E}(\hat{\sigma}_{n, 0})=\frac{2}{n(n-1)}\mbox{tr}(\Sigma^2)+\frac{4}{n(n+1)}\mu_0^{\prime}\Sigma \mu_0.$$ Then from the Markov inequality, we obtain $\hat{\sigma}_{n, 0}/\sigma_n =O_p(1)$ as $p \to \infty$. This indicates that the power $B_1(||\mu_0||^2)$ is largely determined by the signal-to-noise ratio  
\[
\mbox{SNR}_1=\frac{\sqrt{n(n-1)}||\mu_0||^2}{\sqrt{2\mbox{tr}(\Sigma^2)+4(n-1)\mu_0^{\prime}\Sigma \mu_0}}.  
\]
A direct observation shows that $B_1(||\mu_0||^2) \to 1$ if $\mbox{SNR}_1 \to \infty$ as $p \to \infty$. However, the test may lose its power when $\mu_0$ is sparse and weak. To appreciate this, we consider $\Sigma=I_p$ which is the $p \times p$ identity matrix, and let $p^{1-\beta}$ be the number of non-zero components in $\mu_0$ and $\delta$ be the value of each non-zero component. Based on the setup, we observe that when the sample size $n$ is finite, $\mbox{SNR}_1=o(1)$ if $\beta > 1/2$ and $\delta^2=o(p^{\beta-1/2})$, which implies that the proposed test loses its power for sparse and weak signals. On the other hand, when the sample size $n$ diverges, $\mbox{SNR}_1=o(1)$ if $\beta > 1/2$ and $\delta^2=o(n^{-1}p^{\beta-1/2})$, where  the minimum $\delta^2$ required to avoid power loss is $n$ times less than that with a finite $n$.

\section{Two-sample test}
\label{sec:two-sample}

Let $i=1$ or $2$ and $X_{i1}, \cdots, X_{in_i}$ be two independent and identically distributed $p$-dimensional random samples with mean $\mu_i$ and covariance matrix $\Sigma_i$. The two-sample testing problem considers the hypotheses
\begin{eqnarray}
H_0: \mu_1=\mu_2 \quad \mbox{versus} \quad H_1: \mu_1 \ne \mu_2. \label{Hypo2}
\end{eqnarray}
 
If the two samples have the same sample size $n$, we can directly extend the one-sample statistic (\ref{stat1}) to the two-sample case by replacing $X_i$ with $X_{1i}-X_{2i}$ for $i=1, \cdots, n$. However,  two sample sizes are different in many cases. Without loss of generality, we assume $n_1 \le n_2$ and consider 
\begin{eqnarray}
Y_i=X_{1i}-\sqrt{\frac{n_1}{n_2}}X_{2i}+\frac{1}{\sqrt{n_1 n_2}} \sum_{j=1}^{n_1} X_{2j}-\frac{1}{n_2} \sum_{j=1}^{n_2} X_{2j}, \quad i=1, \cdots, n_1. \label{sample.dif}
\end{eqnarray}
This procedure for the difference of two samples was first suggested by Scheffe (1943) in the univariate case to construct the confidence intervals by using the $t$-distribution. It was extended by Anderson (2003) in the multivariate case to obtain the generalized Hotelling's $T^2$ statistic. We adopt the same procedure to propose the two-sample statistic     
\begin{eqnarray}
V_{n_1 n_2}=\frac{2}{n_1(n_1-1)}\sum_{i < j}^{n_1}Y_{i}^{\prime}Y_{j}, \label{stat2}
\end{eqnarray}
which is similar to (\ref{stat1}) but we replace $X_i$ with $Y_i$. 

Note that $\bar{Y}=n_1^{-1}\sum_{i=1}^{n_1}Y_i=n_1^{-1}\sum_{i=1}^{n_1}X_{1i}-n_2^{-1}\sum_{i=1}^{n_2}X_{2i}$. We can write 
\[
V_{n_1 n_2}=\frac{n_1}{n_1-1}\bar{Y}^{\prime} \bar{Y}- \frac{1}{n_1(n_1-1)}\sum_{i=1}^{n_1} Y_i^{\prime} Y_i,
\]
where the first term on the right hand side uses the difference of the two sample means $n_1^{-1}\sum_{i=1}^{n_1}X_{1i}-n_2^{-1}\sum_{i=1}^{n_2}X_{2i}$, which is the most relevant to $\mu_1-\mu_2$. We subtract the second term from the first term so that $\mbox{E}(V_{n_1 n_2})= (\mu_1-\mu_2)^{\prime}(\mu_1-\mu_2)$.
%one advantage of using $Y_i$ is that all the information from the two samples has been utilized for the difference of the two population means.  

Similar to (\ref{model}), we model the two independent and identically distributed $p$-dimensional random samples by the linear high-dimensional time series
\begin{eqnarray}
X_{ij}=\mu_i+\Gamma_i Z_{ij} \qquad \mbox{for} \quad i=1 \quad \mbox{and} \quad 2,  j=1, \cdots, n_i, \label{model2}
\end{eqnarray}
where $\Gamma_i$ is a $p \times q_i$ matrix with $q_i\ge p$ satisfying $\Gamma_i \Gamma_i^{\prime}=\Sigma_i$, and $Z_{ij}=(z_{ij1}, \cdots, z_{ijq_i})^{\prime}$ so that $\{z_{ijl}\}_{l=1}^{q_i}$ are mutually independent and satisfy $\mbox{E}(z_{ijl})=0$, $\mbox{Var}(z_{ijl})=1$ and $\mbox{E}(z_{ijl}^4)=3+\eta$ for some finite constant $\eta$.

By analogy with (C1), we consider the following condition for the two covariance matrices $\Sigma_1$ and $\Sigma_2$.

\medskip
(C2). As $p \to \infty$, $\mbox{tr}(\Sigma_i \Sigma_j \Sigma_k \Sigma_h)=o[\mbox{tr}^2\{(\Sigma_1+\Sigma_2)^2\}]$ for $i, j, k, l=1$ or $2$. 

Under $H_0$ of (\ref{Hypo2}), the variance of $V_{n_1,n_2}$ is
\begin{eqnarray}
\sigma^2_{{n_1 n_2},0}&=&\frac{2}{n_1(n_1-1)}\sigma^2_{{Y^{\prime}Y},0},\label{var20}
\end{eqnarray} 
where $$\sigma^2_{{Y^{\prime}Y},0}=\mbox{tr}(\Sigma_1^2)+\frac{n_1^2}{n_2^2}\mbox{tr}(\Sigma_2^2)+\frac{2n_1}{n_2}\mbox{tr}(\Sigma_1 \Sigma_2)$$ is the variance of $Y_{i}^{\prime}Y_j$ for $i<j$. Similar to the proof of Theorem1, $\{Y_{i}^{\prime}Y_j, i<j\}_{i, j=1}^{n_1}$ can be shown to be a sequence of $n_1(n_1-1)/2$ independent random variables under $H_0$ of (\ref{Hypo2}). We therefore estimate $\sigma^2_{{Y^{\prime}Y},0}$ by 
\begin{eqnarray}
\hat{\sigma}^2_{{Y^{\prime}Y},0}&=& \frac{1}{n_1(n_1-1)/2-1} \sum_{i < j}^{n_1} \biggl(Y_{i}^{\prime}
Y_{j}-\frac{2}{n_1(n_1-1)}\sum_{i<j}^{n_1}Y_{i}^{\prime}
Y_{j}\biggr)^2,\label{var20.est}
\end{eqnarray} 
which is similar to (\ref{var.est}) but we replace $n$ and $X_i$ by $n_1$ and $Y_i$, respectively.  

As a result, the unbiased estimator of $\sigma^2_{{n_1 n_2},0}$ is 
\begin{eqnarray}
\hat{\sigma}^2_{{n_1 n_2},0}&=&\frac{2}{n_1(n_1-1)}\hat{\sigma}^2_{{Y^{\prime}Y},0}.\nonumber
\end{eqnarray} 

The following theorem establishes the asymptotic $t$-distribution of $V_{n_1n_2}/\hat{\sigma}_{n_1 n_2, 0}$, which is a direct extension of Theorem 3. 

\medskip
{\bf Theorem 4.} Assume the model (\ref{model2}) and the condition (C2). For any finite sample sizes $n_1 \ge 3$ and $n_2 \ge n_1$ and under $H_0$ of (\ref{Hypo2}), as $p \to \infty$,  
\[
\frac{V_{n_1 n_2}}{\hat{\sigma}_{n_1 n_2, 0}} \xrightarrow{d} t \quad \mbox{with} \quad k=n_1(n_1-1)/2-1 \quad \mbox{degrees of freedom}.
\]
\medskip
Based on Theorem 4, the proposed test with a nominal $\alpha$ level of significance rejects $H_0$ if $V_{n_1 n_2}/\hat{\sigma}_{n_1 n_2, 0} \ge t_{\alpha}(k_2)$, where $t_{\alpha}(k_2)$ is the upper $\alpha$ quantile of $t$-distribution with $k=n_1(n_1-1)/2-1$ degrees of freedom. Moreover, as $p \to \infty$, the power of the two-sample test is
  \begin{eqnarray}
B_2(||\mu_1-\mu_2||^2)
&=&1-\Phi\biggl(\frac{\hat{\sigma}_{n_1 n_2, 0}}{\sigma_{n_1 n_2}} t_{\alpha}(k)-\mbox{SNR}_2\biggr),
 \end{eqnarray}
where the signal-to-noise ratio 
\[
\mbox{SNR}_2=\frac{\sqrt{n_1(n_1-1)}||\mu_1-\mu_2||^2}{\sqrt{2\mbox{tr}\{(\Sigma_1+\frac{n_1}{n_2} \Sigma_2)^2\}+4(n_1-1)(\mu_1-\mu_2)^{\prime}(\Sigma_1+\frac{n_1}{n_2} \Sigma_2)(\mu_1-\mu_2)}}.
\]
Similar to the one-sample test, the power of the two-sample test is largely determined by $\mbox{SNR}_2$, the analysis of which demonstrates that the proposed test is powerful in detecting dense and strong differences between $\mu_1$ and $\mu_2$, but encounters a power loss when differences between $\mu_1$ and $\mu_2$ are sparse and weak.

\section{Simulation studies}
\label{sec:simulation}
\subsection{One-sample test}

We compare the proposed one-sample $t$-test with the one-sample version CQ test in Chen and Qin (2010), the BS test in Bai and Sarandasa (1996), and
the SD test in Srivastava and Du (2008).  
%In the main paper, we demonstrated the results with data simulated from $N(\mu, \Sigma)$. The results with non-Gaussian data were included in the supplementary material.  

\begin{table}[t]
\tabcolsep 3pt
\centering
\caption{Empirical sizes of Chen and Qin's test (CQ), Bai and Sarandasa's test (BS), Srivastava and Du's test (SD), and the proposed test (New), based on $1000$ replications with normally distributed $Z_i$ in (\ref{model}).}
\label{table1}
\begin{tabular}{c cccc c cccc c cccc}
\hline  &\multicolumn{4}{c}{$p=200$}& &\multicolumn{4}{c}{$p=400$} &&\multicolumn{4}{c}{$p=1000$}  \\[1mm]
\cline{2-5} \cline{7-10} \cline{12-15} & $n=4$& $6$& $15$& $30$&& $n=4$& $6$ &$15$& $30$&& $n=4$& $6$ &$15$ &$30$  \\
\hline
\multicolumn{15}{c}{Model (a)}\\\hline
CQ & $0.129$ & $0.088$& $0.055$ &$0.058$ && $0.122$& $0.086$ &$0.062$ & $0.061$&& $0.134$ & $0.078$& $0.062$ & $0.052$ \\[1 mm]
BS & $0.088$ & $0.080$& $0.056$& $0.056$& & $0.090$& $0.084$ &$0.061$ & $0.064$& &$0.095$& $0.059$ & $0.059$ & $0.053$ \\[1 mm]
SD    & $0.217$    & $0.149$    & $0.055$   & $0.047$  &  & $0.205$  & $0.118$   & $0.046$    & $0.038$  &  & $0.213$  &$0.072$ & $0.018$ & $0.024$    \\[1 mm]
New & $0.056$ & $0.062$& $0.051$ &$0.055$ && $0.059$& $0.059$ &$0.058$ & $0.058$&& $0.058$ & $0.052$ & $0.059$ & $0.050$ 
\\[1 mm]
\hline
\multicolumn{15}{c}{Model (b)}\\\hline
CQ & $0.138$ & $0.079$& $0.044$ &$0.057$ && $0.112$& $0.080$ &$0.067$ & $0.059$&& $0.119$ & $0.070$& $0.050$ & $0.057$ \\[1 mm]
BS & $0.099$ & $0.069$& $0.047$& $0.057$& & $0.064$& $0.079$ &$0.062$ & $0.060$& &$0.090$& $0.061$ & $0.049$ & $0.058$ \\[1 mm]
SD    & $0.218$    & $0.157$    & $0.043$   & $0.035$  &  & $0.192$  & $0.114$   & $0.042$    & $0.036$  &  & $0.227$  &$0.101$ & $0.014$ & $0.033$    \\[1 mm]
New & $0.067$ & $0.049$& $0.042$ &$0.057$ && $0.040$& $0.063$ &$0.064$ & $0.059$&& $0.046$ & $0.053$ & $0.048$ & $0.055$ 
\\[1 mm]
\hline
\end{tabular}
\end{table}

\begin{table}[t]
\tabcolsep 3pt
\centering
\caption{Empirical sizes of Chen and Qin's test (CQ), Bai and Sarandasa's test (BS), Srivastava and Du's test (SD), and the proposed test (New), based on $1000$ replications with $t$ distributed $Z_i$ in (\ref{model}).}
\label{table1-2}
\begin{tabular}{c cccc c cccc c cccc}
\hline  &\multicolumn{4}{c}{$p=200$}& &\multicolumn{4}{c}{$p=400$} &&\multicolumn{4}{c}{$p=1000$}  \\[1mm]
\cline{2-5} \cline{7-10} \cline{12-15} & $n=4$& $6$& $15$& $30$&& $n=4$& $6$ &$15$& $30$&& $n=4$& $6$ &$15$ &$30$  \\
\hline
\multicolumn{15}{c}{Model (a)}\\\hline
CQ & $0.145$ & $0.076$& $0.066$ &$0.062$ && $0.130$& $0.074$ &$0.054$ & $0.060$&& $0.116$ & $0.089$& $0.054$ & $0.056$ \\[1 mm]
BS & $0.072$ & $0.043$& $0.047$& $0.042$& & $0.053$& $0.034$ &$0.032$ & $0.044$& &$0.040$& $0.038$ & $0.028$ & $0.037$ \\[1 mm]
SD    & $0.193$    & $0.089$    & $0.033$   & $0.032$  &  & $0.187$  & $0.070$   & $0.018$    & $0.031$  &  & $0.175$  &$0.047$ & $0.007$ & $0.014$    \\[1 mm]
New & $0.064$ & $0.053$& $0.061$ &$0.059$ && $0.048$& $0.049$ &$0.051$ & $0.058$&& $0.049$ & $0.065$ & $0.051$ & $0.054$ 
\\[1 mm]
\hline
\multicolumn{15}{c}{Model (b)}\\\hline
CQ & $0.143$ & $0.085$& $0.067$ &$0.052$ && $0.141$& $0.069$ &$0.076$ & $0.044$&& $0.128$ & $0.064$& $0.050$ & $0.069$ \\[1 mm]
BS & $0.051$ & $0.039$& $0.038$& $0.039$& & $0.062$& $0.033$ &$0.048$ & $0.025$& &$0.049$& $0.023$ & $0.021$ & $0.049$ \\[1 mm]
SD    & $0.157$    & $0.106$    & $0.038$   & $0.028$  &  & $0.169$  & $0.080$   & $0.030$    & $0.021$  &  & $0.148$  &$0.034$ & $0.006$ & $0.017$    \\[1 mm]
New & $0.056$ & $0.058$& $0.061$ &$0.052$ && $0.068$& $0.049$ &$0.068$ & $0.043$&& $0.059$ & $0.046$ & $0.048$ & $0.069$ 
\\[1 mm]
\hline
\end{tabular}
\end{table}

To generate random samples, we considered two types of innovations in (\ref{model}): the Gaussian $Z_i \sim N(0, I_p)$ and the standardized $t$-distribution with $4$ degrees of freedom for each component of $Z_i$, where the latter has heavier tails than the former used to demonstrate nonparametric performance of the proposed test.  
Under $H_0$, we simply assumed $\mu=0$. Under $H_1$,  we considered $\mu$ to have $[p^{1-\beta}]$ non-zero entires which were randomly selected from $\{1,\cdots, p\}$. Here $[a ]$ denotes the integer part of $a$. The value of each non-zero entry was $r$. From the simulation setup, the two parameters $\beta>0$ and $r>0$ were chosen to control the sparsity and strength of signals, respectively. 
We also considered 
the following two structures for the covariance $\Sigma$, where model (a) specifies a bandable structure of $\Sigma$ and Model (b) leads to a sparse $\Sigma$. 
\begin{description}
\item[(a).]  AR(1) model:  $\sigma_{j_1j_2}=0.6^{|j_1-j_2|}$ for $1\le j_1, j_2\le p$.
\item[(b).]  Random sparse matrix model: first generate a $p \times p$ matrix $\Gamma$ each row of which has only four non-zero element that is randomly chosen from $\{1, \cdots, p\}$ with magnitude generated from Unif(1, 2) multiplied by a random sign. Then $\Sigma=\Gamma \Gamma^{T} + I_{p}$ where $I_{p}$ is the $p \times p$ identity matrix.
\end{description}
All the simulation results were based on $1000$ replications with the nominal significance level $\alpha=0.05$.

\begin{table}[t!]
\tabcolsep 5pt
\centering
\caption{Empirical powers of Chen and Qin's test (CQ), Bai and Sarandasa's test (BS), Srivastava and Du's test (SD), and the proposed test (New), based on $1000$ replications.}
\label{table2}
\begin{tabular}{c cccc cc cccc}
\hline \multicolumn{11}{c}{Normally distributed $Z_i$ with $p=400, \, n=30, \, \beta=0.4$}\\
\hline  &\multicolumn{4}{c}{Model (a)}& & &\multicolumn{4}{c}{Model (b)}\\[1mm]
\cline{2-5} \cline{8-11} & $r=0.1$& $0.2$& $0.3$& $0.4$&& & $r=0.1$& $0.2$ &$0.3$& $0.4$  \\
\hline
CQ & $0.099$ & $0.292$& $0.709$ &$0.990$ && &$0.095$& $0.310$ &$0.796$ & $0.999$ \\[1 mm]
BS & $0.098$ & $0.290$& $0.711$& $0.990$& && $0.094$& $0.311$ &$0.795$ & $0.999$ \\[1 mm]
SD    & $0.062$    & $0.211$    & $0.596$   & $0.972$  & & & $0.058$  & $0.237$   & $0.726$    & $0.991$      \\[1 mm]
New & $0.097$ & $0.288$& $0.705$ &$0.991$ &&& $0.094$& $0.307$ &$0.794$ & $0.999$ 
\\[1 mm]
\hline \multicolumn{11}{c}{$t$ distributed $Z_i$ with $p=400, \, n=30, \, \beta=0.4$}\\
\hline  &\multicolumn{4}{c}{Model (a)}& & &\multicolumn{4}{c}{Model (b)}\\[1mm]
\cline{2-5} \cline{8-11} & $r=0.1$& $0.2$& $0.3$& $0.4$&& & $r=0.1$& $0.2$ &$0.3$& $0.4$  \\
\hline
CQ & $0.097$ & $0.290$& $0.728$ &$0.985$ && &$0.093$& $0.296$ &$0.792$ & $0.996$ \\[1 mm]
BS & $0.068$ & $0.241$& $0.650$& $0.953$& && $0.060$& $0.233$ &$0.728$ & $0.978$ \\[1 mm]
SD    & $0.039$    & $0.170$    & $0.535$   & $0.906$  & & & $0.033$  & $0.175$   & $0.672$    & $0.986$      \\[1 mm]
New & $0.096$ & $0.287$& $0.724$ &$0.985$ &&& $0.090$& $0.290$ &$0.792$ & $0.996$ 
\\[1 mm]
\hline
\end{tabular}
\end{table}

Table \ref{table1} displays the empirical sizes of the four tests with normally distributed $Z_i$ in (\ref{model}) under models (a) and (b) for the covariance matrix $\Sigma$, respectively. The sample size and dimension were chosen to be $n=4, 6, 15, 30$ and $p= 200, 400, 1000$. %Since the size is not affected by the neighborhood size as long as $k=o(n)$, we chose $k=3$ for the NEAT test under the null hypothesis. 
While the CQ, BS and SD tests were able to maintain the empirical sizes close to the nominal significance level $\alpha=0.05$ when sample sizes were relatively large, they encountered size distortion especially when sample size was very small ($n=4$). Unlike the competitors, the proposed test always had the empirical sizes close to the nominal significance level. Similar results can be observed in Table \ref{table1-2} with $t$ distributed $Z_i$ in (\ref{model}). The results confirm Theorem 3 that the proposed testing procedure was established without requiring the diverging sample size and without assuming Gaussian distribution of data.

Due to the size distortion of the CQ, BS and SD tests when sample sizes are small, we compared the power performance of the four tests with a relatively large sample size $n=30$.  
Table \ref{table2} demonstrates the empirical powers of the four tests with respect to different signal strength $r$ when the sparsity of signal $\beta=0.4$. 
As we can see, the powers of the four tests were increased as the signal strength $r$ increased. The proposed test performed similarly to the other three tests. This is not surprising as the four tests were all proposed based on similar sum-of-squares type statistics.

\begin{figure}[t!]
\centering
\includegraphics[width=0.49\textwidth,height=0.5\textwidth]{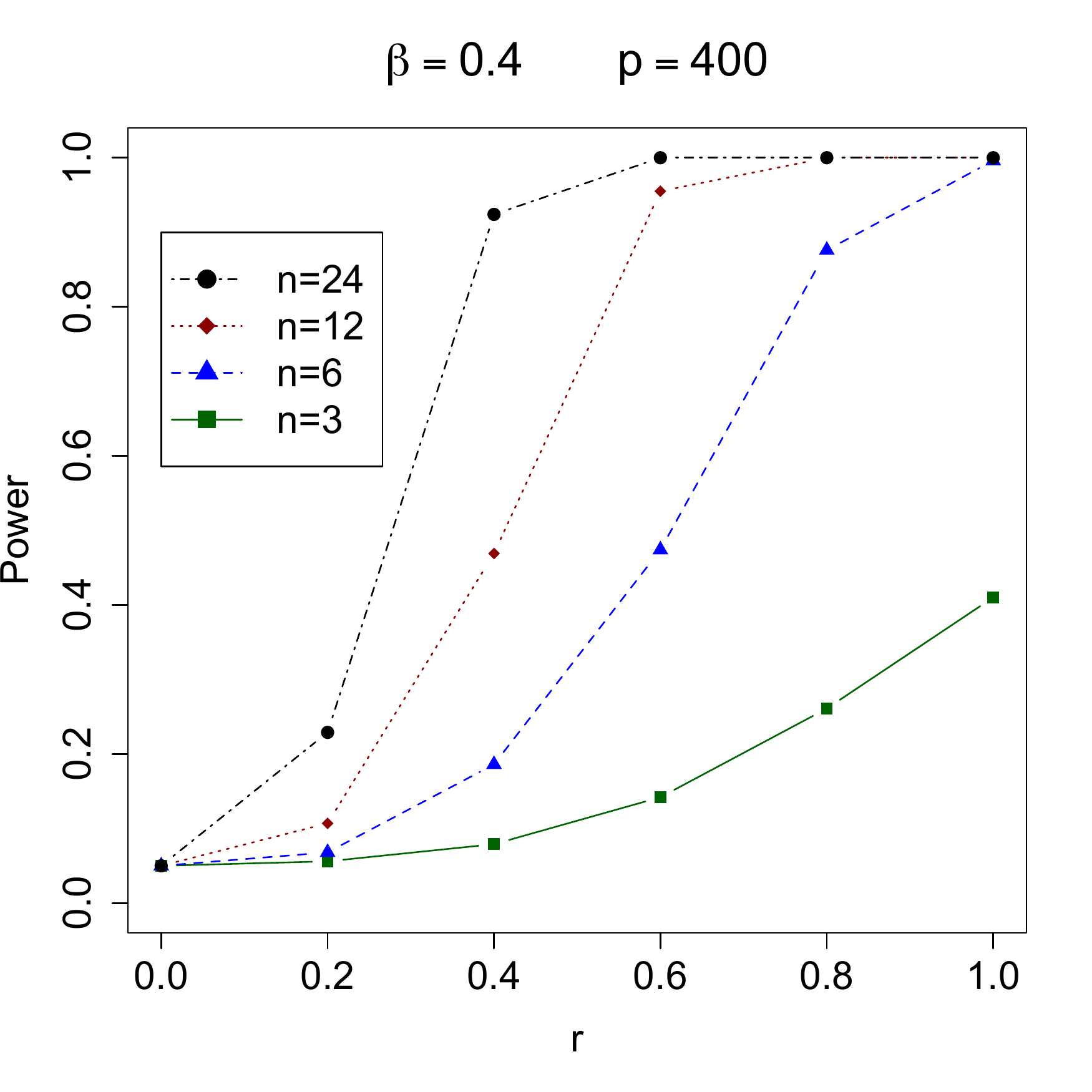}
\includegraphics[width=0.49\textwidth,height=0.5\textwidth]{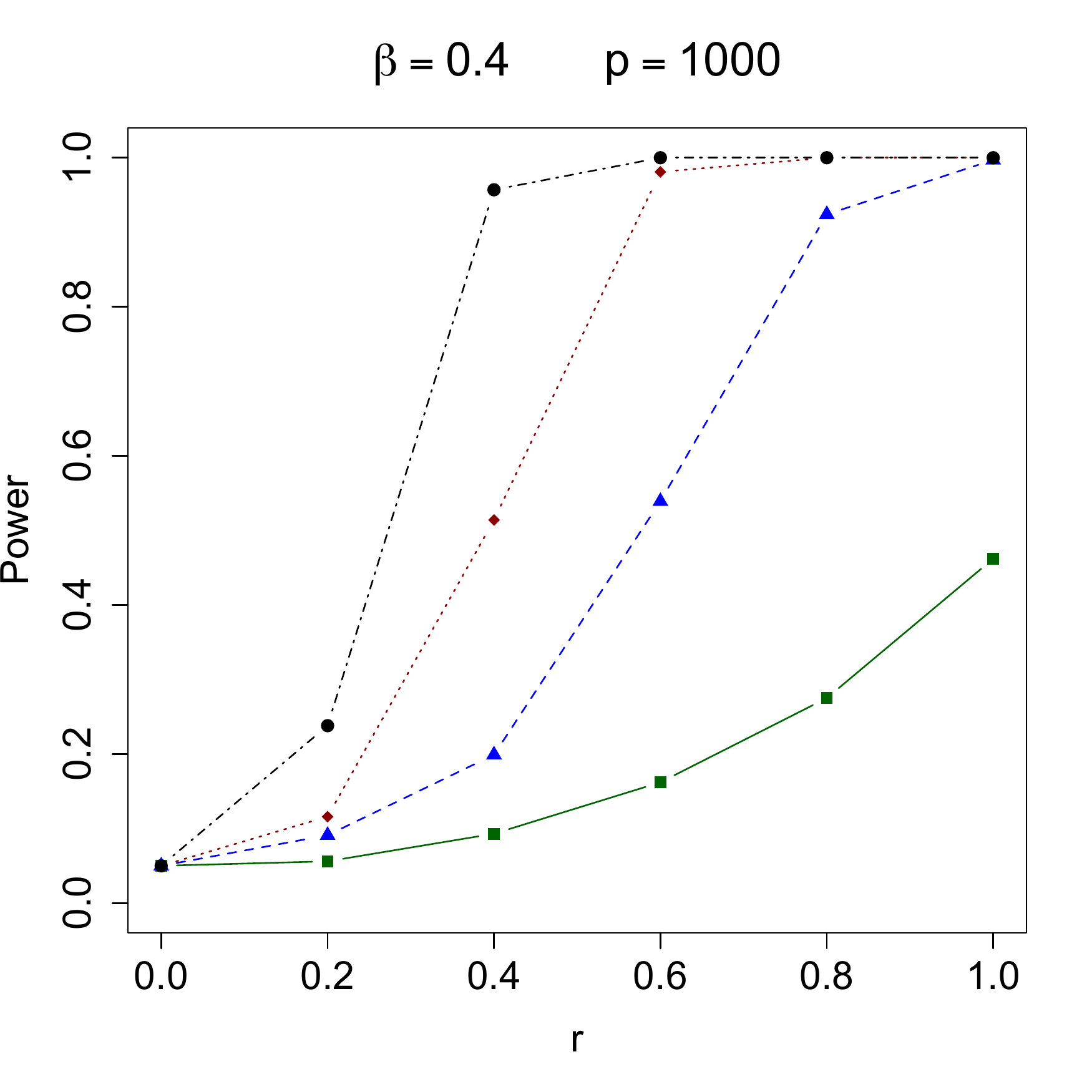}\\
  \caption{Empirical powers of the proposed test with respect to signal strength $r$ by choosing different sample sizes and dimensionalities.}
  \label{fig1}
\end{figure}

To further investigate the power performance of the proposed test with small sample sizes, we chose a range of sample sizes $n=3, 6, 12, 24$ with normally distributed $Z_i$. For each of the sample sizes, the empirical powers of the proposed test were obtained with respect to a range of signal strength $r$ from $0.1$ to $1$. As illustrated in Figure \ref{fig1}, the powers increased as the signal strength $r$ increased for each sample size $n$, and as the sample size increased for each signal strength $r$. Compared the right panel with $p=1000$ to the left panel with $p=400$, the same pattern was observed but the powers  were greater when dimension $p$ became larger. %Similar pattern were observed with $t$ distributed $Z_i$. Due to space limitation, we omit the results.      

\subsection{Two-sample test}

\begin{table}[t!]
\tabcolsep 2pt
\centering
\caption{Empirical sizes of Chen and Qin's test (CQ), Cai, Liu and Xia's test (CLX), Chen, Li and Zhong's test (CLZ), and the proposed test (New), based on $1000$ replications. The sample size $n_2=30$, and the sample size $n_1$ increased from $4$ to $30$. }
\label{table3}
\begin{tabular}{c cccc c cccc c cccc}
\hline \multicolumn{15}{c}{Normally distributed $Z_{1j}$ and $Z_{2j}$}\\
\hline  &\multicolumn{4}{c}{$p=200$}& &\multicolumn{4}{c}{$p=400$} &&\multicolumn{4}{c}{$p=1000$}  \\[1mm]
\cline{2-5} \cline{7-10} \cline{12-15} $n_2=30$ & $n_1=4$& $6$& $15$& $30$&& $n_1=4$& $6$& $15$& $30$&& $n_1=4$& $6$& $15$& $30$  \\
\hline
CQ & $0.108$ & $0.084$& $0.064$ &$0.068$ && $0.085$& $0.069$ &$0.060$ & $0.060$&& $0.106$ & $0.067$& $0.059$ & $0.054$ \\[1 mm]
CLX & $0.971$ & $0.599$& $0.046$& $0.014$& & $0.994$& $0.763$ &$0.062$ & $0.024$& &$1$& $0.934$ & $0.076$ & $0.003$ \\[1 mm]
CLZ    & $0.998$    & $0.869$    & $0.113$   & $0.036$  &  & $1$  & $0.980$   & $0.168$    & $0.025$  &  & $1$  &$1$ & $0.236$ & $0.015$    \\[1 mm]
New & $0.054$ & $0.062$& $0.055$ &$0.066$ && $0.058$& $0.053$ &$0.059$ & $0.059$&& $0.057$ & $0.060$ & $0.055$ & $0.061$ 
\\[1 mm]
\hline \multicolumn{15}{c}{$t$-distributed $Z_{1j}$ and $Z_{2j}$}\\
\hline  &\multicolumn{4}{c}{$p=200$}& &\multicolumn{4}{c}{$p=400$} &&\multicolumn{4}{c}{$p=1000$}  \\[1mm]
\cline{2-5} \cline{7-10} \cline{12-15} $n_2=30$ & $n_1=4$& $6$& $15$& $30$&& $n_1=4$& $6$& $15$& $30$&& $n_1=4$& $6$& $15$& $30$  \\
\hline
CQ & $0.095$ & $0.054$& $0.052$ &$0.060$ && $0.096$& $0.060$ &$0.063$ & $0.062$&& $0.089$ & $0.066$& $0.054$ & $0.051$ \\[1 mm]
CLX & $0.968$ & $0.582$& $0.060$& $0.006$& & $0.996$& $0.745$ &$0.058$ & $0.008$& &$1$& $0.929$ & $0.081$ & $0.001$ \\[1 mm]
CLZ    & $0.999$    & $0.849$    & $0.137$   & $0.034$  &  & $1$  & $0.973$   & $0.166$    & $0.019$  &  & $1$  &$1$ & $0.266$ & $0.011$    \\[1 mm]
New & $0.055$ & $0.050$& $0.054$ &$0.061$ && $0.056$& $0.048$ &$0.065$ & $0.061$&& $0.049$ & $0.059$ & $0.063$ & $0.050$ 
\\[1 mm]
\hline
\end{tabular}
\end{table}

Under $H_0$, we compared the size performance of the proposed test with the two-sample version CQ test in Chen and Qin (2010), the maximum type CLX test in Cai, Liu and Xia (2014), and the multi-level thresholding CLZ test in Chen, Li and Zhong (2019). The random samples were generated from two types of innovations in (\ref{model2}). The Gaussian $Z_{1j} \sim N(0, I_p)$ and $Z_{2j} \sim N(0, I_p)$, and the standardized $t$-distribution with $4$ degrees of freedom for each component of $Z_{1j}$ and $Z_{2j}$. For simplicity, we assigned $\mu_1=\mu_2=0$, and considered $\Sigma_1=\Sigma_2$ modeled by the AR(1) structure (a) in Section 4.1. %Simulation studies with non-Gaussian data were also considered and the results were included in the supplementary material.  

Table \ref{table3} displays the empirical sizes of the four tests. %Similar to the one-sample case, we chose $k=3$ for the NEAT test under the null hypothesis. 
The sample size $n_2=30$, and the sample size $n_1$ increased from $4$ to $30$. 
The dimensions of random vector were $p= 200, 400$ and $1000$.
For all the cases, the proposed test maintained the empirical sizes close to the nominal significance level $\alpha=0.05$. However, the CQ, CLX and CLZ tests had inflated sizes especially when the sample size $n_1$ was small at $4$.

%Under $H_1$, $\mu_1=0$ but $\mu_2$ had $[p^{1-\beta}]$ non-zero entires which were randomly selected from $\{1,\cdots, p\}$. The value of each non-zero entry was $r$.  All the simulation results were based on $1000$ replications with nominal significance level $\alpha=0.05$.
Under $H_1$, the power of the proposed test performed as well as the CQ test when sample sizes were relatively large. The pattern of the empirical results was very similar to the one-sample case given by Table \ref{table2}. We therefore omit the results. It is worth mentioning that when the two high dimensional mean vectors differ only in sparsely coordinates and the differences are faint, the CLX and CLZ tests were proposed to improve power performance of the CQ test. It is therefore not surprising that they perform better than the proposed test for sparse and faint signal detection. But we need to emphasize that such a superior performance relies on the requirement of relatively large sample sizes. %Despite this, the proposed test outperforms the CQ, CLX and CLZ tests when sample sizes are very small, because it was established by only requiring dimension to diverge.          

\section{Application to fMRI dataset}
\label{sec:application}

To demonstrate the practical use of the proposed tests, we consider the StarPlus fMRI data, which is publicly available from Carnegie Mellon University’s Center for Cognitive Brain Imaging. The original data consist of different trials and we use a subset in which each of six human subjects was provided a sentence first for four seconds, followed by a blank screen for four seconds. The subject was then provided a picture for four seconds, followed by answering whether the sentence correctly described the picture. At last, the subject was given a rest for fifteen seconds. There are in total $55$ images  collected every $0.5$ seconds. At each time point, the image is marked with 25-30 anatomically defined regions called regions of interest (ROIs)(Mitchell et al. 2003). In fMRI, ROI analysis is a useful method of selecting a cluster of voxels for exploring patterns of activation across stimuli.

\begin{table}[t!]
\tabcolsep 3pt
\centering
\caption{The 15 ROIs in the StarPlus dataset tested by the proposed methods.}
\label{table4}
\begin{tabular}{|c|c|}
\hline 
\multicolumn{1}{|c|}{ROI}& \multicolumn{1}{c|}{Full Name} \\\hline
CALC & Calcarine Sulcus\\\hline
LDLPFC/RDLPFC &Left/Right Dorsolateral Prefrontal Cortex  \\\hline
LIPL &  Left Inferior Parietal Lobule \\\hline
LIPS/RIPS &Left/Right Intraparietal Sulcus \\\hline
LIT/RIT &Left/Right Inferior Temporal Lobe\\\hline
LOPER/ROBER &Left/Right Opercularis \\\hline
LSPL/RSPL & Left Superior Parietal Lobe \\\hline
LT/RT & Left/Right Temporal Lobe \\\hline
LTRIA/RTRIA & Left/Right Triangularis \\\hline
SMA & Supplementary Motor Area\\
\hline
\end{tabular}
\end{table}

\begin{figure}[t!]
\centering
\includegraphics[width=0.6\textwidth,height=0.6\textwidth]{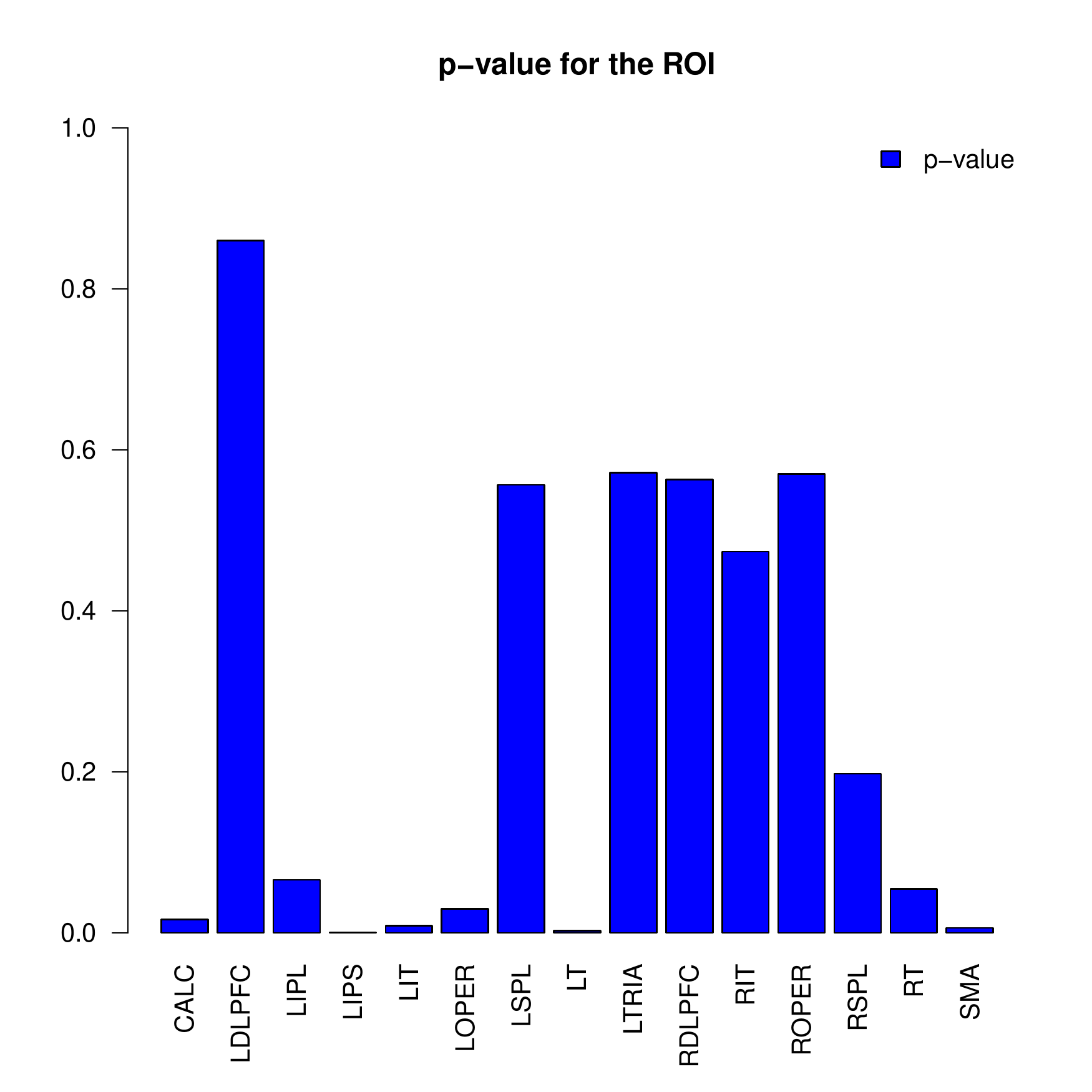}
  \caption{Bar plot of the p-values for the 15 ROIs considered by the proposed test.}
  \label{fig2}
\end{figure}

Our interest is to identify the ROIs which react differently to a sentence and a picture. To accommodate high dimensionality, we consider the ROIs with the number of voxels greater than 60. 
The names of these ROIs are described in Table \ref{table4}. 
We let $\mu_{1i}$ and $\mu_{2i}$ be the population means of the $i$th ROI with respect to a picture and a sentence, respectively. The null hypotheses of interest are $H_{0i}: \mu_{1i}=\mu_{2i}$ for $i=1, \cdots, 15$, where $15$ is the number of ROIs. Since the dataset has a very sample size $6$, it is more appropriate to apply the proposed test rather than other competitors requiring diverging sample sizes. 
For each ROI, we computed the difference between the fMRI image at 29 seconds and that at 9 seconds. The two time points are the ends of the time intervals during which the picture and the sentence were presented, respectively. We applied the proposed test to obtain the p-values of the $15$ ROIs displayed by Figure \ref{fig2}. By further applying the Bonferroni correction to control the family-wise error rate at $0.05$, the two ROIs  
named LIPS and LT were identified to be significant as their p-values were less than 0.05/15.

%\section*{\large 6. \bf Conclusion and discussion}

%For the problem of detecting a change point in the mean, we propose a max-type procedure and a sum-type procedure. The proposed procedures assume the population mean before and after the change point, and the covariance matrix across components of a random vector are all unknown. To study theoretical properties of the proposed procedures, we derive their average run lengths (ARLs) and expected detection delays (EDDs). The numerical studies confirm the accuracy of the derived ARLs, and suggest a preference for the proposed max-type procedure.  

%While we assume the random vectors are normally distributed, the proposed procedures are expected to work for non-Gaussian data as well. Similar to Li and Li (2019), more technical conditions are needed to establish the theoretical results when data are non normally distributed. The proposed max-type and sum-type procedures in the high dimensional setting are the analog of the CUSUM and Shiryayev-Roberts procedures, respectively, in the univariate setting. The previous numerical studies show that the CUSUM procedure outperforms the Shiryayev-Roberts procedure in terms of the supremum conditional expected detection delay. Our numerical investigation achieves the same conclusion for the proposed max-type and sum-type procedures. %Except the supremum conditional expected detection delay, there exist other evaluation criteria. It would be interesting to compare the two procedures under different criteria.     

\section{Appendix: Technical Details.}

\bigskip
\noindent{\bf A.1. Proof of Theorem 1.}
\bigskip

From $\{X_i, 1 \le i \le n\}$, we construct a sequence of $n(n-1)/2$ random variables $\{X_i^{\prime} X_j, i<j\}_{i,j=1}^n$. To establish the asymptotic normality of $U_n$, we need to show that the sequence $\{X_i^{\prime}X_j/\sqrt{\mbox{tr}(\Sigma^2)+2\mu^{\prime}\Sigma \mu}, i<j\}_{i, j=1}^n$ converges to a joint multivariate normal distribution as $p \to \infty$. According to the Cramer-word device, we only need to show that $\sum_{i < j}^{n}c_{ij}X_i^{\prime}X_j/\sqrt{\mbox{tr}(\Sigma^2)+2\mu^{\prime}\Sigma \mu}$ is asymptotically normal as $p \to \infty$, where $\{c_{ij}, i<j\}_{i, j=1}^n$ are some constants and at least one of them is nonzero. 

We first establish the asymptotic normality of $\sum_{i < j}^{n}c_{ij}X_i^{\prime} X_j/\sqrt{\mbox{tr}(\Sigma^2)}$ under the null hypothesis. From (\ref{model}), we see that $X_i=\Gamma Z_i$. Based on that,
\begin{eqnarray}
\sum_{i < j}^{n}\frac{c_{ij}X_i^{\prime} X_j}{\sqrt{\mbox{tr}(\Sigma^2)}}&=&\sum_{i<j}^n \frac{c_{ij}}{\sqrt{\mbox{tr}(\Sigma^2)}} Z_i^{\prime} \Gamma^{\prime}\Gamma  Z_j
= \sum_{k=1}^p \sum_{l=1}^p \sum_{i<j}^n \frac{c_{ij}}{\sqrt{\mbox{tr}(\Sigma^2)}} (\Gamma^{\prime}\Gamma)_{kl}z_{ik} z_{jl}. \nonumber
\end{eqnarray}

To simplify notation, we define
\[
A_{ij,kl}=\frac{c_{ij}}{\sqrt{\mbox{tr}(\Sigma^2)}} (\Gamma^{\prime}\Gamma)_{kl},
\]
if $k > l$, and if $k=l$, 
\[
A_{ij,kk}=\frac{ c_{ij}}{2\sqrt{\mbox{tr}(\Sigma^2)}} (\Gamma^{\prime}\Gamma)_{kk}.
\]
Then, using the symmetry, we can write 
\begin{eqnarray}
\sum_{i < j}^{n}\frac{c_{ij}X_i^{\prime} X_j}{\sqrt{\mbox{tr}(\Sigma^2)}}&=&\sum_{k=1}^p \sum_{l=1}^k \sum_{i < j}^{n}A_{ij, kl }(z_{ik} z_{jl}+z_{il} z_{jk})=\sum_{k=1}^p V_k, \nonumber
\end{eqnarray}
where $V_k=\sum_{l=1}^k \sum_{i < j}^{n}A_{ij, kl }(z_{ik} z_{jl}+z_{il} z_{jk})$. Let $S_h=\sum_{k=1}^h V_k$. Since $\mbox{E}(S_q|S_h)=S_h$ for any $q > h$, we see that $S_h$ is a martingale. We therefore use the Martingale central limit theorem to establish the asymptotic normality of $\sum_{i < j}^{n}{c_{ij}X_i^{\prime} X_j}/{\sqrt{\mbox{tr}(\Sigma^2)}}$. According to the Martingale central limit theorem, it is equivalent to proving the following two results:
\begin{eqnarray}
{\sum_{k=1}^p \mbox{E}(V_k^2| \mathcal{F}_{k-1})} \xrightarrow{p} \sum_{i<j}^n c_{ij}^2, \quad \mbox{and} \label{le1}
\end{eqnarray}
\begin{eqnarray}
{\sum_{k=1}^p \mbox{E}\{V_k^2\mbox{I}(|V_k|>\epsilon | \mathcal{F}_{k-1}\}} \xrightarrow{p} 0, \label{le2}
\end{eqnarray}
where $\mathcal{F}_{k-1}$ is the $\sigma$ algebra generated by $\{z_{i1}, \cdots, z_{ik-1}\}$ for $i=1, \cdots,n$, and $\epsilon$ is any small positive number.

To prove (\ref{le1}), we need to show $\sum_{k=1}^p \mbox{E}(V_k^2) \to \sum_{i<j}^n c_{ij}^2$ and $\mbox{Var}\{\sum_k \mbox{E}(V_k^2| \mathcal{F}_{k-1})\} \to 0$, respectively. To this end, we notice
\[
V_k^2=\sum_{l_1=1}^k \sum_{l_2=1}^k \sum_{i_1<j_1}^n \sum_{i_2<j_2}^n A_{i_1j_1,k l_1}A_{i_2j_2,k l_2} (z_{i_1k}z_{j_1l_1}+z_{i_1l_1}z_{j_1k})(z_{i_2k}z_{j_2l_2}+z_{i_2l_2}z_{j_2 k}).
\]
Then, 
\[
\sum_{k=1}^p \mbox{E}(V_k^2)=2\sum_{k=1}^p \sum_{l=1}^k\sum_{i<j}^n A_{ij,k l}^2+2\sum_{k=1}^p\sum_{i<j}^n A_{ij,k k}^2=\sum_{i<j}^n c_{ij}^2. 
\]
 
Next, it can be seen that
\begin{eqnarray}
\sum_k \mbox{E}(V_k^2| \mathcal{F}_{k-1})&=& 4\sum_{k=1}^p \sum_{i<j} A_{ij,kk}^2+\sum_{k=1}^p\sum_{l_1=1}^{k-1}\sum_{l_2=1}^{k-1}\sum_{i<j_1}\sum_{i<j_2}A_{ij_1,kl_1}A_{ij_2,kl_2}z_{j_1 l_1}z_{j_2 l_2}\nonumber\\
&+&\sum_{k=1}^p\sum_{l_1=1}^{k-1}\sum_{l_2=1}^{k-1}\sum_{i_1<j_1}\sum_{i_2<i_1}A_{i_1j_1,kl_1}A_{i_2i_1,kl_2}z_{j_1 l_1}z_{i_2 l_2}\nonumber\\
&+&\sum_{k=1}^p\sum_{l_1=1}^{k-1}\sum_{l_2=1}^{k-1}\sum_{i_1<i_2}\sum_{i_2<j_2}A_{i_1i_2,kl_1}A_{i_2j_2,kl_2}z_{i_1 l_1}z_{j_2 l_2}\nonumber\\
&+&\sum_{k=1}^p\sum_{l_1=1}^{k-1}\sum_{l_2=1}^{k-1}\sum_{i_1<j_1}\sum_{i_2<j_1}A_{i_1j_1,kl_1}A_{i_2j_1,kl_2}z_{i_1 l_1}z_{i_2 l_2}.\nonumber
\end{eqnarray}
Taking the expectation of the above, we can show that
\begin{eqnarray}
\mbox{E}[\{\sum_k \mbox{E}(V_k^2| \mathcal{F}_{k-1})\}]^2&=&(\sum_{i<j}^n c_{ij}^2)^2+O\biggl\{(\sum_{i<j}^n c_{ij}^2)^2\frac{\mbox{tr}(\Sigma^4)}{\mbox{tr}^2(\Sigma^2)}\biggr\}.\nonumber
\end{eqnarray}
As a result, when $p \to \infty$,  
\begin{eqnarray}
\mbox{Var}\{\sum_k \mbox{E}(V_k^2| \mathcal{F}_{k-1})\}&=&O\biggl\{(\sum_{i<j}^n c_{ij}^2)^2\frac{\mbox{tr}(\Sigma^4)}{\mbox{tr}^2(\Sigma^2)}\biggr\} \to 0,\nonumber
\end{eqnarray}
because $\mbox{tr}(\Sigma^4)=o\{\mbox{tr}^2(\Sigma^2)\}$ according to the condition (C1).

By Chebyshev Inequality, to prove (\ref{le2}), we only need to show that
\[
{\sum_{k=1}^p\mbox{E}(V_k^4)} \to 0.
\]
Using $V_k=\sum_{l=1}^k \sum_{i < j}^{n}A_{ij, kl }(z_{ik} z_{jl}+z_{il} z_{jk})$, we can show that for some constant $C$,
\begin{eqnarray}
\sum_{k=1}^p\mbox{E}(V_k^4)%&=&\sum_{l=2}^q\{15+6\beta-4 \mbox{E}(z_l^6)+\mbox{E}(z_l^8) \}\tilde{G}_{ll}^4 +\sum_{l=2}^q\{-28-12 \beta\nonumber\\
%&+&6 \mbox{E}(z_l^6)\}\tilde{G}_{ll}^2(\sum_{k=1}^{l-1}\tilde{G}_{kl}^2)
%+\sum_{l=2}^q\{3(3+\beta)\} \sum_{k_1 \ne k_2}^{l-1}\tilde{G}_{k_1l}^2 \tilde{G}_{k_2l}^2 \nonumber\\
%&=&C\sum_{l=2}^q(\sum_{k=1}^l \tilde{G}_{kl}^2)^2 \nonumber\\
&\le& \frac{C}{\mbox{tr}^2(\Sigma^2)} (\sum_{i<j} c_{i j})^4  \sum_k \sum_{l_1 l_2}(\Gamma^{\prime} \Gamma)_{kl_1}(\Gamma^{\prime} \Gamma)_{kl_1}(\Gamma^{\prime} \Gamma)_{kl_2}(\Gamma^{\prime} \Gamma)_{kl_2}\nonumber\\
&\le& \frac{C}{\mbox{tr}^2(\Sigma^2)} (\sum_{i<j} c_{i j})^4 \mbox{tr}(\Sigma^4) \to 0,\nonumber
\end{eqnarray}
because $\mbox{tr}(\Sigma^4)=o\{\mbox{tr}^2(\Sigma^2)\}$ according to the condition (C1). 

Based on (\ref{le1}) and (\ref{le2}), we apply the Martingale central limit theorem to establish the asymptotic normality of $\sum_{i < j}^{n}c_{ij}X_i^{\prime} X_j/\sqrt{\mbox{tr}(\Sigma^2)}$ under the null hypothesis.

To prove asymptotic normality of $\sum_{i < j}^{n}c_{ij}X_i^{\prime} X_j$ under the alternative, we use (\ref{model}) to write
\begin{eqnarray}
\sum_{i < j}^{n}c_{ij}X_i^{\prime} X_j&=&\sum_{i<j}^n c_{ij} \mu^{\prime} \mu +\sum_{i<j}^n c_{ij} \mu^{\prime} \Gamma Z_j+\sum_{i<j}^n c_{ij} \mu^{\prime} \Gamma Z_i
+ \sum_{i<j}^n c_{ij} Z_i^{\prime} \Gamma^{\prime} \Gamma Z_j, \nonumber
\end{eqnarray}
where the last term remains under the null hypothesis and its asymptotic normality has been established.  Next, we need to establish the asymptotic normality of $\sum_{i<j}^n c_{ij} \mu^{\prime} \Gamma Z_j+\sum_{i<j}^n c_{ij} \mu^{\prime} \Gamma Z_i$. By observing that $Z_i=(z_{i1}, \cdots, z_{ip})^{\prime}$ and $\{z_k\}_{k=1}^p$ are mutually independent, the asymptotic normality of $\sum_{i<j}^n c_{ij} \mu^{\prime} \Gamma Z_j+\sum_{i<j}^n c_{ij} \mu^{\prime} \Gamma Z_i$ can be established from the Lyapunov's condition. At last, $\sum_{i<j}^n c_{ij} \mu^{\prime} \Gamma Z_j+\sum_{i<j}^n c_{ij} \mu^{\prime} \Gamma Z_i$ and $\sum_{i<j}^n c_{ij} Z_i^{\prime} \Gamma^{\prime} \Gamma Z_j$ are asymptotically independent. We thus establish the asymptotic normality of $\sum_{i < j}^{n}c_{ij}X_i^{\prime} X_j$ under the alternative hypothesis. This completes the proof of Theorem 1.

\bigskip
\noindent{\bf A.2. Proof of Theorem 2.}
\bigskip

In Theorem 1, we have shown that under the null hypothesis, for $1\le i <j \le n$, $(X_1^{\prime} X_2, \cdots, X_i^{\prime} X_j, \cdots, X_{n-1}^{\prime} X_n)^{\prime}$  follows an asymptotic $n(n-1)/2$-variate multivariate normal distribution with mean equal to zero and covariance matrix equal to $\mbox{tr}(\Sigma^2)I_{n(n-1)/2}$, where $I_{n(n-1)/2}$ is the $n(n-1)/2 \times n(n-1)/2$ identity matrix. Based on $\{X_i^{\prime} X_j, i<j\}_{i,j=1}^n$, we estimate the unknown $\mbox{tr}(\Sigma^2)$ by the sample variance (\ref{var.est}). Since $\{X_i^{\prime} X_j, i<j\}_{i,j=1}^n$ are asymptotically independent and normally distributed random variables, ${k \,\, \widehat{\mbox{tr}({\Sigma}^2)}}/{\mbox{tr}(\Sigma^2)}$ converges to $\chi^2(k)$  as 
$p \to \infty$, where $k=n(n-1)/2-1$. This completes the proof of Theorem 2.    

\bigskip
\noindent{\bf A.3. Proof of Theorem 3.}
\bigskip

The statistic $U_n$ is the sample mean of $\{X_i^{\prime} X_j, i<j\}_{i,j=1}^n$. From the proof of Theorem 2, $U_n$ is asymptotically independent of the sample variance (\ref{var.est}). As a result, 
${U_n}/{\hat{\sigma}_{n, 0}}$ converges to a $t$-distribution with  $k=n(n-1)/2-1$ degrees of freedom. This completes the proof of Theorem 3. 

\bigskip
\noindent{\bf A.4. Proof of Theorem 4.}
\bigskip

Theorem 4 can be shown by replacing $X_i$ with $Y_i$ in the proofs of Theorems 1-3. We therefore omit it.

\section*{Reference}

\begin{description}

\item
{Anderson, T. W.} (2003), \textit{An Introduction to Multivariate Statistical Analysis}, Hoboken, NJ: Wiley.

%\textsc{Ashburner, M., Ball, C.A., Blake, J.A., Botstein, D., Butler, H., Cherry, J.M., Davis, A., Dolinski, K., Dwight, S.S., Eppig, J.T., et al.} (2000), ``The Gene Ontology Consortium. Gene Ontology: tool for the unification of biology'', \textit{ Nature Genetics}, 25, 25-29.

%\textsc{Bai, Z.} (1993), ``Convergence Rate of Expected Spectral Distributions of Large Random Matrices. Part II. Sample Covariance Matrices'', \textit{The Annals of Probability}, 21, 649-672.

%\textsc{Bai, Z., Yin, Y. Q.} (1993), ``Limit of the Smallest Eigenvalue of Large Dimensional Covariance Matrix'', \textit{The Annals of Probability}, 21, 1275-1294.

%\textsc{Bai, Z., Jiang, D., Yao, J. and Zheng, S.} (2009), ``Corrections to LRT on Large-Dimensional Covariance Matrix by RMT'', \textit{The Annals of Statistics}, 37, 3822-3840.
\item
{Bai, Z., and Saranadasa, H.} (1996),
``Effect of High Dimension: By an Example
of a Two Sample Problem'', \textit{Statistica Sinica}, 6, 311-329.

%\textsc{Bai, Z. and Silverstein, J.} (2010), \textit{Spectral Analysis of Large Dimensional Random Matrices}, Springer.
%\item
%{Barry, W., Nobel, A., and Wright, F.} (2005), ``Significance Analysis of Functional Categories in Gene Expression Studies: A Structured Permutation Approach'', \textit{Bioinformatics}, 21, 1943-1949.

%\textsc{Benjamini, Y., and Hochberg, Y.} (1995), ``Controlling the False Discovery Rate: A Practical and Powerful Approach to Multiple Testing'', \textit{Journal of the Royal Statistical Society Series B}, 57, 289-300.
%\item
%{Bickel, P. J. and Levina, E.} (2008a), ``Regularized Estimation of Large Covariance Matrices'', \textit{The Annals of Statistics}, 36, 199-227.
%\item
%{Bickel, P. J. and Levina, E.} (2008b), ``Covariance Regularization by Thresholding'', \textit{The Annals of Statistics}, 36, 2577-2604.

%\textsc{Cai, T., Liu, W. and Luo, X.} (2011), ``A Constrained $l_1$ Minimization Approach to Sparse Precision Matrix Estimation'', \textit{Journal of the American Statistical Association}, 106, 594-607.

\item
{Cai, T., Liu, W. and Xia, Y.} (2014), ``Two-Sample Test of High Dimensional Means under Dependence'', \textit{Journal of the Royal Statistical Society: Series B}, 76, 349-372.
%\item
%{Cai, T., Zhang, C. and Zhou, H.} (2010), ``Optimal Rates of Convergence for Covariance Matrix Estimation'', \textit{The Annals of Statistics}, 38, 2118-2144.

\item
{Chen, S. X., Li, J. and Zhong, P.} (2019),
`` Two-Sample and ANOVA Tests for High Dimensional Means'', \textit{The Annals of Statistics}, 47, 1443-1474.
%\item
%{Chen, L., Paul, D., Prentice R. and Wang, P.} (2011),
%``A Regularized Hotelling's T2 Test for Pathway Analysis in Proteomic Studies'', \textit{Journal of the American Statistical Association}, 106, 1345-1360.
\item
{Chen, S. X., and Qin, Y.-L.} (2010),
``A Two Sample Test for High Dimensional
Data With Applications to Gene-Set Testing'', \textit{The Annals of Statistics}, 38,
808-835.

\item
{Hall, P., and Jin, J.} (2010), ``Innovated Higher Criticism for Detecting Sparse Signals in Correlated Noise'', \textit{The Annals of Statistics}, 38, 1686-1732.

\item
{Hotelling, H.} (1931), ``The Generalization of Student’s ratio'', \textit{Annals of Mathematical Statistics}, 2, 360-378.
\item
{Li, J., and Chen, S. X.} (2012), ``Two Sample Tests for High Dimensional Covariance Matrices'', \textit{The Annals of Statistics}, 40, 908-940.

\item
{Mitchell, T., Hutchinson, R., Just, M. A., Niculescu, R., Pereira, F. and Wang, X.} (2003), ``Classifying Instantaneous Cognitive States from fMRI Data'', \textit{The American Medical Informatics Association 2003 Annual Symposium}, 465-469.

\item
\textsc{Scheffe, H.} (1943), ``On Solutions of the Behrens-Fisher Problem, based on the $t$-distribution'', \textit{Annals of Mathematical Statistics}, 14, 35-44.

\item
{Srivastava, M.S. and Du, M.} (2008), ``A Test for the Mean Vector with Fewer Observations than the Dimension", \textit{Journal of Multivariate Analysis}, 99, 386-402.
\item
{Srivastava, R., Li, P. and Ruppert, D.} (2016), ``RAPTT: An Exact Two-Sample Test in High Dimensions Using Random Projections", \textit{Journal of Computational and Graphical Statistics}, 25, 954-970.

\item
{Thulin, M.} (2014), ``A High-Dimensional Two-Sample Test for the Mean Using Random Subspaces'', \textit{Computational Statistics and Data Analysis}, 74, 26-38.

%\textsc{van der Laan, M., and Bryan, J.} (2001), ``Gene Expression Analysis With the Parametric Bootstrap'', \textit{Biostatistics}, 2, 445-461.
%\item
%{Wagaman, A. S. and Levina, E.} (2009), ``Discovering Sparse Covariance Structures with the Isomap", \textit{Journal of Computational and Graphical Statistics}, 18, 551-572.
%\item
%{Wang, L., Peng, B. and Li, R.} (2015), ``A High-Dimensional Nonparametric Multivariate Test for Mean Vector", \textit{Journal of the American Statistical Association}, 110, 1658-1669. 
\item
{Zhong, P., Chen, S. X. and Xu, M.} (2013), ``Tests Alternative to Higher Criticism for High Dimensional Means under Sparsity and Column-wise Dependence'', \textit{The Annals of Statistics}, 41, 2820-2851.\end{description}

\end{document}